# Waveguide Excitation and Spin Pumping of Chirally Coupled Quantum Dots

Savvas Germanis,[1,†] Xuchao Chen,[1,†] René Dost,[1] Dominic J. Hallett,[1] Edmund Clarke,[2] Pallavi K. Patil,[2] Maurice S. Skolnick,[1] Luke R. Wilson,[1] Hamidreza Siampour,[3,*] A. Mark Fox[1,*]

[1] School of Mathematical and Physical Sciences, University of Sheffield, Hicks Building, Sheffield S3 7RH, UK

[2] EPSRC National Epitaxy Facility, School of Electrical and Electronic Engineering, University of Sheffield, Sheffield S1 3JD, UK

[3] School of Mathematics and Physics, Queen's University Belfast, University Road, Belfast BT7 1NN, UK

[*] Emails: h.siampour@qub.ac.uk, mark.fox@sheffield.ac.uk

**Abstract**

We report on an integrated semiconductor chip where a single quantum dot (QD) is excited in-plane via a photonic-crystal waveguide through its nearest p-shell optical transition. The chirality of the waveguide mode is exploited to achieve both directional absorption and directional emission, resulting in a substantial enhancement in directional contrast, as measured for the Zeeman components of the waveguide-coupled QD. This remote excitation scheme enables high directionality ($\geq 0.95$) across ~56% of the waveguide area, with significant overlap with the Purcell-enhanced region, where the electric field intensity profile is near its peak. In contrast, local excitation methods using an out-of-plane excitation beam focused directly over the area of the QD achieve only ~25% overlap. This enhancement increases the likelihood of locating Purcell-enhanced QDs in regions that support high directionality, enabling the experimental demonstration of a six-fold enhancement in the decay rate of a QD with >90% directionality. The remote p-shell excitation protocol establishes a new benchmark for waveguide quantum optics in terms of the combination of Purcell enhancement and high directionality, thereby paving the way for on-chip excitation of spin-based solid-state quantum technologies in regimes of high β-factor.

**Introduction**

The rapid evolution of photonic quantum technologies is driven by potential breakthroughs in quantum computing, communication, and sensing.[1-3] Central to this progress is the development of chip-scale quantum optical circuits that seamlessly integrate the excitation, manipulation, and detection of single photons.[4-8] A key component of this integration is on-chip excitation, where quantum emitters are excited via waveguide modes rather than using an off-chip laser. This technique is essential for implementing precise quantum operations, miniaturising devices, and enabling more compact quantum technologies.[5,9] For instance, recent advancements in quantum sensing and light guiding, such as the use of fluorescent nanodiamond-doped polyvinyl alcohol (PVA) fibres, demonstrate promising applications in on-chip excitation.[10] This method offers several advantages, including reduced optical crosstalk, minimised local heating, the ability to excite multiple emitters using a single mode, and access to emitters in hard-to-reach regions of a photonic device. Additionally, this technique can be extended to chiral quantum systems, where engineering and preserving the polarization and flow of light within the waveguide allows information stored in polarized states to be transmitted unidirectionally.[11]

---

[†] S.G. and X.C. contributed equally to this work.

A key parameter for waveguide quantum optics is the β-factor, which quantifies the probability that an emitted or scattered photon is coupled to the waveguide mode. Values of the β-factor close to unity are required to observe strong quantum-optical effects such as single-photon nonlinearities, and the high-β regime acquires additional interest when the quantum emitter is positioned at a chiral point of the waveguide.[11] For such a chirally-coupled quantum emitter, a spin-dependent phase shift of π can be imparted to a photon propagating in a specific direction, and this can be exploited to develop quantum spin-networks.[12] There have been numerous observations of chiral emission from quantum emitters in a variety of nano-photonic systems[13-20], and in a recent paper we demonstrated a notable combination of chiral emission with a high Purcell factor, and hence β-factor.[21] However, the demonstration of directional absorption associated with the chirality of the waveguide modes has received far less attention, with all previous studies restricted to nanobeam waveguides in regimes of only moderate β-factor.[9,22,23] The in-plane excitation geometry permits efficient spin pumping via a remote laser of arbitrary polarization[9], and establishes protocols for spin-dependent dot-to-dot interactions in a chip-based geometry. This capability is crucial for developing spin-based quantum computing architectures and advancing quantum communication technologies. Furthermore, it enables the interaction of quantum spin states from two or more QDs within photonic devices, mediated through photons, which is essential for the implementation of spin cluster states.[12,24,25]

In this work, we push the boundaries of in-plane excitation by exploiting the chirality of slow-light waveguide modes to initialise and manipulate exciton spins in the high β-factor regime. We employ a remote excitation method that significantly enhances the directional contrast under a linearly-polarized pump. Specifically, we demonstrate that, when combined with glide-plane photonic crystal waveguides, in-plane p-shell excitation enables the efficient initialisation of circularly polarized spin states, leading to enhanced directionality of the emitted photons. These photons are coupled to the right- or left-circularly polarized propagating waveguide mode, depending on their polarization. This approach not only improves the directionality of emitter-waveguide interactions but also facilitates the coherent manipulation and readout of quantum information encoded in these spins.

**Remote Excitation Method**

Figure 1 illustrates the device layout and working principle. The device consists of a glide-plane photonic-crystal waveguide containing quantum dots with in- and out-couplers at opposite ends, as illustrated in Fig. 1a. The quantum dots are located within a p-i-n diode as shown in Fig. 1b to enable exciton tuning via the quantum-confined Stark effect. This tunability is particularly important for photonic-crystal waveguides in which slow-light enhancement is only achieved at specific wavelengths. On application of a strong external magnetic field in Faraday geometry, the QD exciton states are circularly polarized, with the Zeeman components having opposite helicities. If the quantum dot is located at a chiral point of the waveguide, the modes are circularly polarized (see inset to Fig. 1a), with the helicity depending on the direction of propagation. Hence photons with opposite circular polarizations generated by exciton recombination propagate to either left or right according to their helicity. The Zeeman lines are therefore observed with differing intensities at the out-couplers, with their intensity ratio depending on the relative populations of the exciton spin states and on the chirality of the waveguide.

Figures 1c and 1d contrasts two possible methods to excite the quantum dot, namely *local* and *remote*. For local excitation, the pump laser is focused directly onto the QD, as shown in Figure 1c. This is the method employed in the great majority of previous studies,[13-15,21] and the intensity ratio of the Zeeman components is assumed to be determined by the chirality of the waveguide mode at the position where the QD is located. For this assumption to be correct, it is necessary that the populations of the $\sigma^+$ and $\sigma^-$ polarized exciton spin states should be identical. The assumption is valid for the case of resonant or near-resonant pumping with a linearly-polarized or unpolarized laser, and for non-resonant pumping with any polarization if there is no spin memory.

Figure 1d illustrates the contrasting case of the remote excitation scheme, where the laser is focussed on one of the grating couplers at the end of the device and the quantum dot is excited via photons propagating within the waveguide. In this case, a quantum dot at a chiral point will be driven by a circularly-polarized field irrespective of the polarization of the pump laser, with the helicity of the circular field depending on the direction of propagation. In our experiment, the laser is tuned to the p-shell of the QD, and initialises the spin of the p-shell exciton states depending on the helicity of the waveguide mode. The p-shell photogenerated carriers relax non-radiatively to form s-shell excitons, preserving the spin memory of their initial state.[26-28] The s-shell excitons recombine radiatively emitting circularly polarized photons into the waveguide. The photons of different helicities then propagate in opposite directions according to the chirality at the QD's position.

The chief difference between the two excitation methods is that the chirality of the waveguide enters twice into the intensity ratio of the Zeeman components for the remote excitation method. First, the initial populations of the $\sigma^+$ and $\sigma^-$ excitons are determined by the helical pumping at the chiral point, and hence are not the same. Second, the directional emission of the opposite spin states is again determined by the chirality. These two effects combine to produce a much stronger directional contrast between the Zeeman components for remote excitation compared to local, where the chirality only affects the results at the second stage. The successful observation of enhanced directionality for remote excitation thus establishes the effectiveness of the chiral spin-pumping method, and hence demonstrates the validity of photon-mediated spin-spin interaction in chip-based photonic circuits.

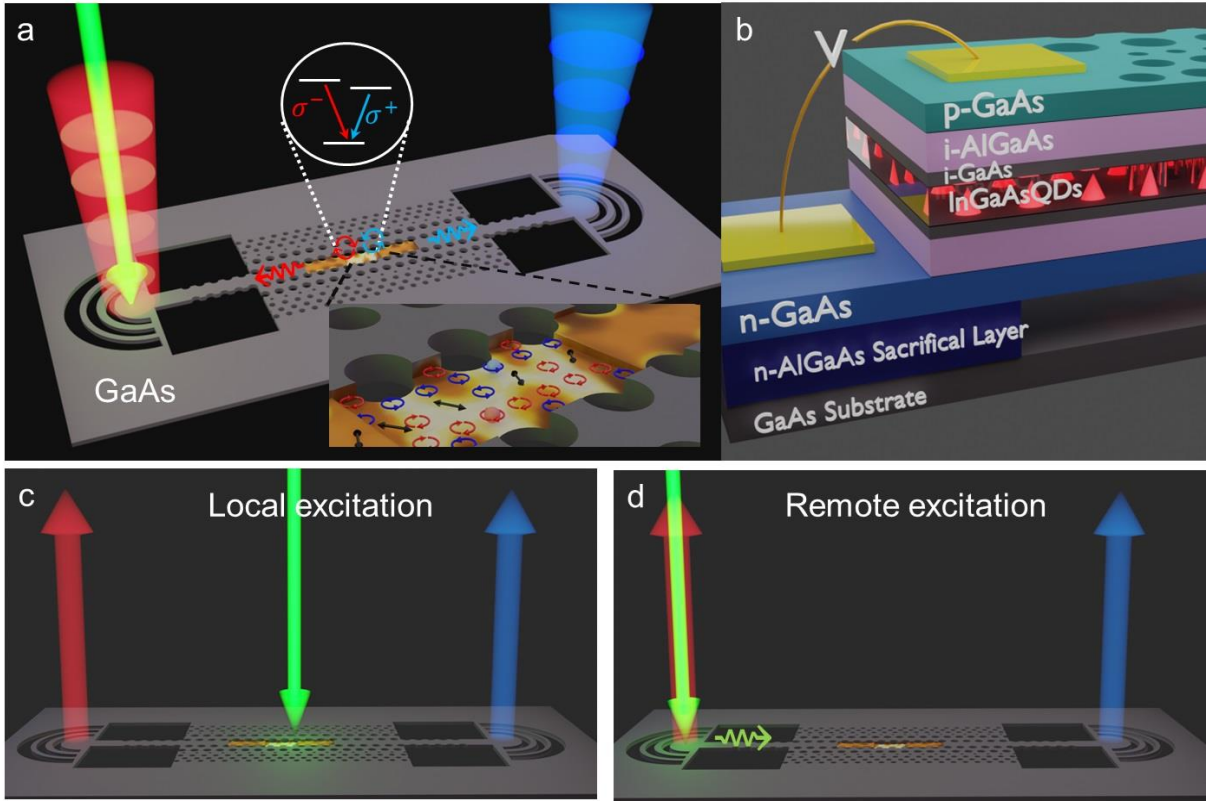

**Figure 1**: Schematic of the device layout and working principle. (**a**) A semiconductor chip where a pump laser, indicated by a green arrow, is coupled into a waveguide using a grating. The waveguide directs light towards a QD positioned within the chiral region of the waveguide. The helicity of the waveguide mode in the photonic-crystal line defect enables selective excitation of the QD spins. Once excited, photons emitted from the QD couple to the waveguide mode and propagate either to the left (red) or right (blue) along the waveguide, depending on their circular polarization. The inset shows the intensity profile and helicity (chirality) of the waveguide mode within the slow-light section of the photonic-crystal line defect. (**b**) Schematic of the p-i-n GaAs diode structure with embedded InGaAs QDs and electrical contacts (yellow) made to the p- and n-GaAs layers. (**c**) Schematic of the local excitation scheme, where the pump laser is focused directly onto the QD. (**d**) Schematic of the remote excitation scheme, where the QD is excited via waveguide modes.

## Numerical Simulations

Figure 2a presents the simulated chirality map of the glide-plane photonic crystal waveguide used in the device's slow-light region. The map employs the Stokes parameters, $S_3 = -2Im(E_x E_y^*)$ and $S_0 = I$, to visualise the degree of circular polarization normalised by light intensity, represented as $C_l = S_3/S_0$. Here $I$ is the intensity, and $E_x$ and $E_y$ are the electric field components within the sample plane. The subscript $l$ indicates that this is the chirality measured using the local excitation method according to the position of the quantum emitter. The map shows strong chirality, with $C_l \geq 0.95$ in certain regions (solid black contour), enabling efficient coupling between exciton spins and the waveguide modes, thus facilitating controlled spin initialisation and manipulation. The waveguide's band structure reveals two bands (Figure 2b), with the simulated Purcell factor shown near the crossing point (i.e. $ka/\pi = 0.44$), where the slow-light effect is most pronounced. The spatial dependence of the chirality and the Purcell factor demonstrate that regions of strong chirality and high Purcell factor coincide. Restricting chirality to $C_l \geq 0.95$ (solid black contour), the overlap region covers more than 25% of the waveguide area, and includes areas with a predicted Purcell factor exceeding 20.

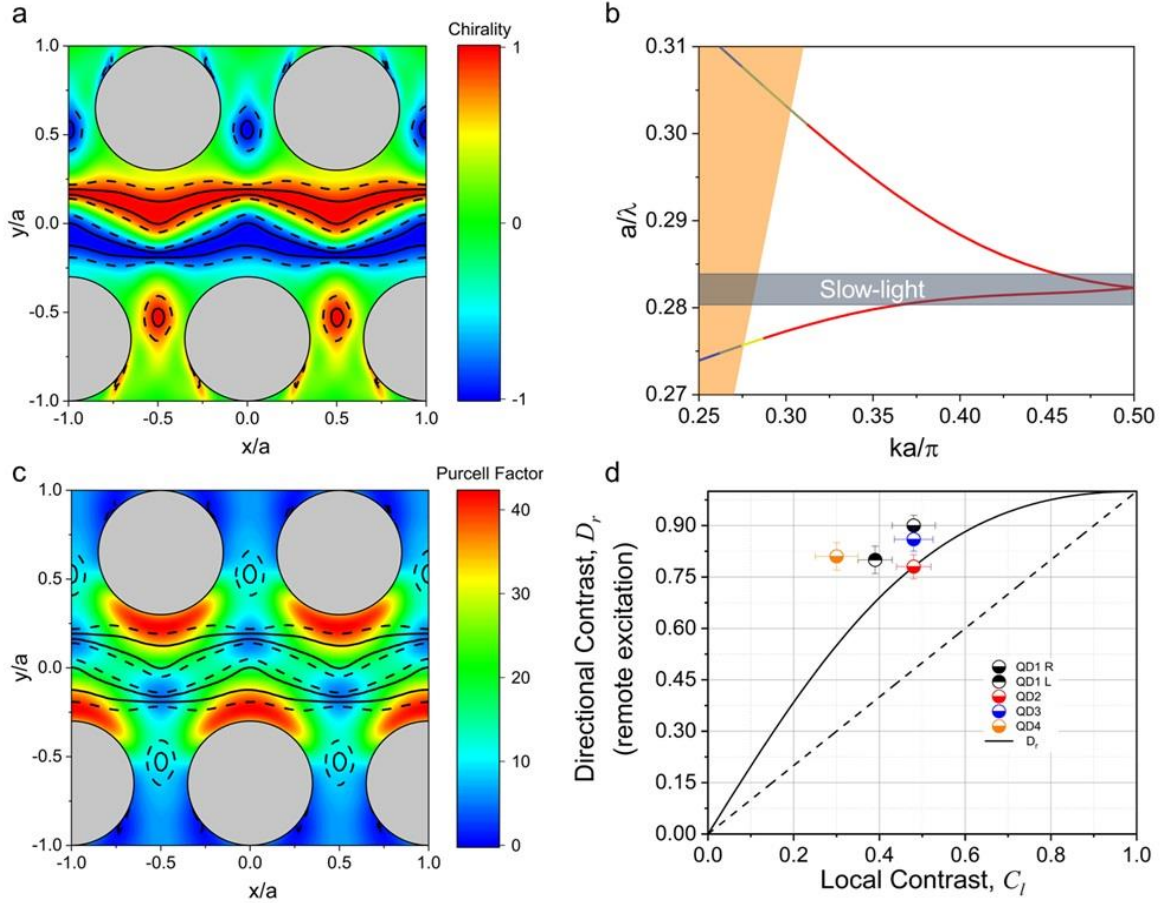

**Figure 2**: **(a)** Simulated chiral map of the photonic-crystal waveguide slow-light region, showing regions of high local chirality ($C_l > 0.95$, solid black contour), calculated using Stokes parameters. These regions enable efficient coupling between QD spins and waveguide modes. Under remote excitation, the high directional contrast ($D_r \geq 0.95$) region covers ~56% of the waveguide area (dashed black contour), compared to ~25% for local excitation. **(b)** Band structure of the waveguide showing two bands, with the Purcell enhancement factor being most pronounced near the crossing point. **(c)** Simulated Purcell factor in the slow-light region. The solid and dashed contours show regions with directional contrast ($\geq 0.95$) under remote and local excitation respectively, as in (a). The solid contour includes areas with Purcell Factor >20, illustrating the efficiency of remote excitation in expanding the highly directional regions for QD spin control. **(d)** The relationship between directional contrast under remote excitation ($D_r$) and the local contrast ($C_l$) exhibits a substantial increase in effective directional contrast for measurements with remote excitation. The solid line is derived from the conversion formula given in Equation (3), the dash line shows $D_r = C_l$ relation and the circles represent experimental data described in the next section.

The helicity of the waveguide mode in the photonic-crystal line defect allows for the selective excitation of QD spins via remote excitation followed by directional absorption (Figure 1d), as opposed to local excitation (Figure 1c). When an incident photon propagating within the waveguide drives a QD dipole transition, the polarization of the local electric field is transferred to the exciton, resulting in the quantum state $|\psi_s\rangle = \alpha_n |\uparrow\rangle + \beta_n |\downarrow\rangle$, where $\alpha$ and $\beta$ are normalised coefficients that satisfy $|\alpha_n|^2 + |\beta_n|^2 = 1$. If $|\alpha_n|^2 \neq |\beta_n|^2$, the imbalance in the population of the two Zeeman states significantly affects the measured directional contrast, as the polarization-dependent absorption is followed by directional emission, and both processes depend on $C_l$. By contrast, for local excitation with linear polarization or in conditions where there is no spin memory, it follows that $|\alpha_n|^2 = |\beta_n|^2$, and the directionality just depends on the value of $C_l$ for the emission process. Consequently, the directional contrast

observed under in-plane remote excitation differs from that observed under local excitation with a linearly polarized pump.

To further analyse this effect, we derive a conversion formula that quantitatively relates the directionality observed through local excitation to that achieved under remote excitation. This relationship holds under the assumptions of the circular dipole approximation, spin-preserving scattering, and weak excitation. Specifically, consider a QD located where the intensity ratio between the Zeeman components via local excitation is $x:1$. The directional contrast in this scenario is given by:

$$C_l = (x-1)/(x+1). \tag{1}$$

If we assume that the populations of the σ⁺ and σ⁻ polarized exciton spin states are identical under local excitation (i.e. $|\alpha_n|^2 = |\beta_n|^2$), then $C_l$ will be equal to the value of $S_3/S_0$ at the position of the quantum dot. This assumption is valid for a linearly polarized pump. It will also be valid for any pump polarization under conditions where the spin memory is lost after excitation.

Under remote excitation, electron-hole pairs are generated with the exciton spin populations reflecting the polarization of the local electric field. In the ideal case of 100% spin fidelity, quantum spin information in the form of photon directionality is fully transferred to the polarization of excitons, with $|\alpha_n|^2 : |\beta_n|^2 = x:1$, or vice versa, depending on the direction of propagation. The excitons then emit with a relative intensity of $x:1$. This spin-selective, directional excitation (absorption), followed by another directional emission process, results in a squared intensity ratio of $x^2:1$ for the emitted photons. Consequently, the effective directional contrast under remote excitation, which reflects the combined effects of both directional absorption and directional emission, is given by:

$$D_r = (x^2-1)/(x^2+1). \tag{2}$$

By substituting $x = (1+C_l)/(1-C_l)$, we obtain the following expression:

$$D_r = 2C_l/(1+C_l^2). \tag{3}$$

This relationship reveals that the directional contrast under remote excitation depends on the initial contrast $C_l$ but exhibits a non-linear enhancement for moderate values of $C_l$, as illustrated in Figure 2d. This suggests that a modest directionality observed under direct excitation using a linearly polarized pump can yield a substantial increase in effective directional contrast for measurements with remote excitation.

For example, if $C_l = 0.5$ (corresponding to an intensity ratio of $3:1$), the resulting effective directional contrast will reach $D_r = 0.8$ – a significant improvement. Additionally, remote excitation can expand the effective area of directionality for waveguide-based photonic devices. In particular, the area with over 95% effective directional contrast increases from 25% to 56% in our glide-plane waveguide devices under remote excitation (dashed black contour in Figure 2a). This represents more than a twofold expansion of the highly directional region compared to local excitation, and the expanded region encompasses areas with strong Purcell enhancement exceeding 20 (solid black contour in Figure 2c). Our remote excitation technique takes advantage of the helicity of the slow-light waveguide mode, enabling the combination of higher Purcell factors and near-unity directional contrast. This balance underscores the potential of remote excitation in optimising chiral photonic systems for waveguide quantum optics.

**Experimental Results**

In the experiment, the sample is housed in a liquid helium cryostat surrounded by superconducting coils, enabling magneto-photoluminescence studies under an applied magnetic field along the sample axis (see setup in Supplementary, Figure S1). For quasi-resonant excitation, the laser is tuned to energies just above the exciton state, specifically targeting the QD's p-shell. As shown in the photoluminescence (PL) excitation spectrum in Figure S2, the p-shell is located at ~896 nm, approximately 20.7 meV (14 nm) above the exciton emission line at 910 nm.

We measured the PL of the exciton emission line under p-shell excitation, while applying an external magnetic field of 3T in the Faraday geometry. For the local excitation scheme (Figure 1c), where the pump laser is focused directly on the QD1, the PL spectra (Figure 3a) exhibit relatively weak directionality. In contrast, under the remote excitation scheme (Figure 1d), where the QD1 is excited via waveguide modes, the PL spectra (Figure 3b) show a significantly enhanced directional contrast. This enhancement is attributed to stronger spin-selective pumping of the QD via the waveguide mode, driven by directional absorption and, in turn, directional coupling of the QD emission within the glide-plane photonic-crystal devices.

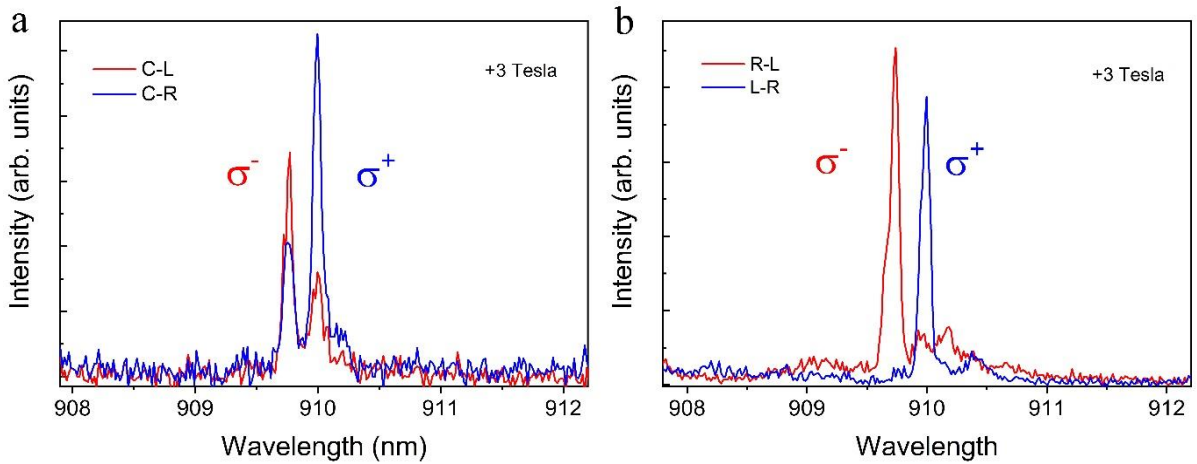

**Figure 3**: PL spectra of the exciton emission line under an external magnetic field of 3T in Faraday geometry where the field is applied along the sample growth axis. **(a)** PL spectra for local excitation and **(b)** corresponding PL spectra for remote excitation. The comparison highlights the significant enhancement in the directional emission under remote excitation, demonstrating improved spin-selective interactions.

In Figure 2d, we present data from a study of four QDs coupled into the glide-plane waveguide, represented by coloured points. The graph compares the directionality for remote excitation and local excitation, with the prediction of Equation (3) for $D_r$ shown by the solid line. The dashed line shows the case where there is no enhancement under remote excitation, i.e. $D_r = C_l$. For all QDs shown in this graph, excitation conditions were carefully selected to ensure p-shell excitation above the QD ground states. Black circular points correspond to the values from QD1, the primary QD studied in this report. The labels $L$ and $R$ indicate the two out-couplers where the PL was collected. The variation between points arises from chiral asymmetry, commonly observed in these devices due to waveguide disorder and multiple scattering processes affecting the propagating mode (see e.g. ref. 13). Red, blue, and orange square points represent PL measurements from QDs in other devices with the same chiral design parameters. These QDs, embedded within the glide-plane photonic-crystal waveguide, exhibit relatively low chiral behaviour under local excitation. In all cases, the directional contrast increases under remote excitation. It is striking that the enhancement is close to, or sometimes larger, than that predicted by Equation (3), which confirms that the spin memory under p-shell excitation is close to 100%. For the results that exceed Equation (3), it could be the case there is some ellipticity in the pump laser, which could result in $|\alpha_n|^2 \neq |\beta_n|^2$ even for local excitation, and hence affect the experimental values of $C_l$.

Figures 4a and 4b present time-resolved PL measurements to assess the decay rate enhancement in the slow-light waveguide for the same QD1 as in Figure 3. In our experiments, electrical and magnetic field tuning were generally employed to achieve a red or blue shift, resulting from either or both the quantum confined Stark effect and Zeeman energy. Specifically, for QD1, electrical field tuning was applied at zero magnetic field to induce a red shift and align the QD's wavelength as closely as possible with the centre of the slow-light band of the photonic-crystal waveguide (Supplementary Figure S4). Figure 4a shows the time-resolved PL at a wavelength of 910.4 nm, where the emission is the fastest. In addition, Figure 4b presents the variation in lifetime with emission wavelength as the electric field is increased and red-shift tuning is applied. It demonstrates a reduction in decay time within the slow-light band as the emission wavelength approaches the centre of the band, followed by an increase in decay time as the wavelength moves beyond the centre, returning to larger values. The single exponential fit in Figure 4a indicates a lifetime of approximately 200 ps, corresponding to a six-fold decay rate enhancement compared to the ensemble lifetime of 1.2 ns for the dots in the wafer. This results in an estimated Purcell factor of 6 for the chirally-coupled QD1, achieving a significant combination of high directional contrast (>90%) and Purcell factor in a QD, and demonstrating spin pumping of the QD1 in the slow light region of glide-plane waveguides where the estimated β-factor, calculated using FDTD simulations, is 97%.

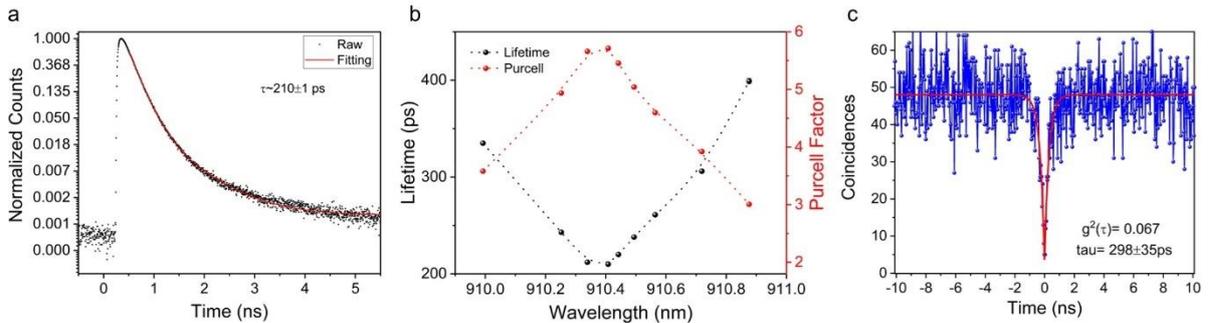

**Figure 4**: Emission characteristics and single-photon verification of the QD. **(a)** Lifetime measurement of the QD at a wavelength of 910.4 nm, demonstrating a lifetime of approximately 200 ps and a six-fold enhancement in the decay rate compared to the ensemble lifetime of 1.2 ns. **(b)** Wavelength dependence of the QD's lifetime and corresponding Purcell factor as the wavelength is tuned by electric field within the slow light region. It shows a decrease in decay time as the emission wavelength nears the center of the slow-light band, followed by an increase as the wavelength moves past the center. **(c)** The second-order correlation function was obtained through HBT correlation measurements on the exciton line under CW laser excitation, with a strong p-shell resonance detuned by ~20.7 meV from the exciton state (Supplementary Figure S2).

Figure 4c shows the results of a Hanbury Brown and Twist (HBT) measurement under continuous-wave (CW) excitation at the p-shell resonance for the same QD1 as in Figure 3. The clear anti-bunching behaviour at zero–time delay ($g^2(0) \sim 0.06$) without background subtraction demonstrates its excellent performance as a chirally-coupled single-photon source in the high-β-factor regime. The HBT results under pulsed excitation are shown in the Supplement (Figure S3).

**Discussion and Conclusions**

By demonstrating enhanced directional contrast through remote excitation and combining it with Purcell enhancement, we introduce a novel mechanism for improving chiral light-matter interactions in nanophotonic platforms in the high-β regime that is required for waveguide quantum optics. The process of spin-photon coupling plays a critical role, that is, the transfer of photon polarization to the exciton spin, followed by its recombination and subsequent re-emission. The method only works when the exciton spin is preserved during relaxation, as any decoherence processes would lower the final chiral response, which is why we use p-shell excitation. In our work we have focussed on exciton spins, but the method can easily be adapted for spin pumping of electrons and holes in charged quantum dots, opening up new possibilities for spin-based quantum networks and quantum information processing systems.

Our investigation demonstrates the potential of remote excitation techniques combined with photonic-crystal waveguides to significantly enhance spin initialisation and directional coupling of QDs. By leveraging the chirality of waveguide modes in the slow-light regime, we have expanded the region of high directionality, theoretically achieving directionality exceeding 95% and a Purcell enhancement greater than 20. The simulations highlight a substantial improvement when using remote excitation, with approximately 56% of the waveguide area having ≥ 95% directionality – compared to only around 25% achieved with local excitation methods. Furthermore, our experimental measurements demonstrate a six-fold enhancement in the emission decay rate of a coupled QD with 90% directional contrast under remote excitation, which corresponds to a β-factor of ~97%, as calculated using FDTD simulations. This advancement enables improved control of quantum states and facilitates the integration of such systems into chip-scale quantum optical circuits. This work provides a foundation for future research to optimise chiral quantum emitter interactions within photonic circuits, an important step toward functional quantum devices.

**Methods**

*Photoluminescence Measurements*: Measurements were performed in a helium bath cryostat (supplementary Figure S1) at T= 4K equipped with a superconducting magnet (0-5 Tesla). The sample was mounted in a socket giving access to electrical control of the PIN devices. The sample holder apparatus was fixed on a X-Y-Z piezo-stage, ensuring stable positioning of the sample. The optical access of the sample was through a confocal scanning microscope set-up. A CW tunable laser (Toptica single-mode laser DL Pro) was used for PL excitation and P-shell excitation experiments, while a femtosecond pulsed Ti:Sapphire laser (Spectra-Physics Tsunami, Newport) with an 80 MHz repetition rate was used for lifetime and PL correlation experiments. Both lasers were fibre-coupled. On the collection path, an ultranarrow bandpass filter with a full width at half maximum (FWHM) of less than 0.55 nm (935.4–0.45 OD5 Ultra Narrow Bandpass Filter, Alluxa) was angle-tuned with respect to the emission line of the QD under study, effectively filtering out unwanted emission lines and the quasi-resonant p-shell excitation laser from the QD signal. PL Spectra were recorded by a liquid nitrogen-cooled charge-coupled device (CCD) camera after being dispersed through a 0.75 Acton Pro monochromator. Time resolved PL measurements were implemented by using a superconducting nanowire fast single-photon detectors (SNSPD - Single Quantum Eos), while the laser pulse repetition rate was detected by a photodiode. The pulses from the SNSPD and the photodiode were analysed using a time-correlated photon counting card (Becker and Hickl SPC-130-EM).

*Simulation:* Numerical calculations were performed using the commercial software package *Lumerical FDTD Solutions* and the open-source Python package *Legume* (see details in the Supplementary section, Section 3).

## Data Availability
The data that support the findings of this study are available from the authors.

## Acknowledgements
This work was funded by the Engineering and Physical Sciences Research Council (EPSRC) UK Programme Grants EP/V026496/1. H.S. acknowledges support from the UKRI Strength in Places Fund programme, Smart Nano NI, and technical assistance from Shelby Hanna in creating Figure 1.

## Author Contributions
S.G. and X.C. performed the measurements. H.S. designed the photonic devices, and R.D. conducted the device fabrication. E.C. and P.K.P. grew the quantum dot wafer. X.C. contributed to the simulations. H.S. and A.M.F supervised the experiment. H.S. wrote the manuscript with input from S.G. and X.C. All authors discussed the results and commented on the manuscript.

# Supplementary

**Section S1**: Cryogenic photoluminescence setup and p-shell excitation

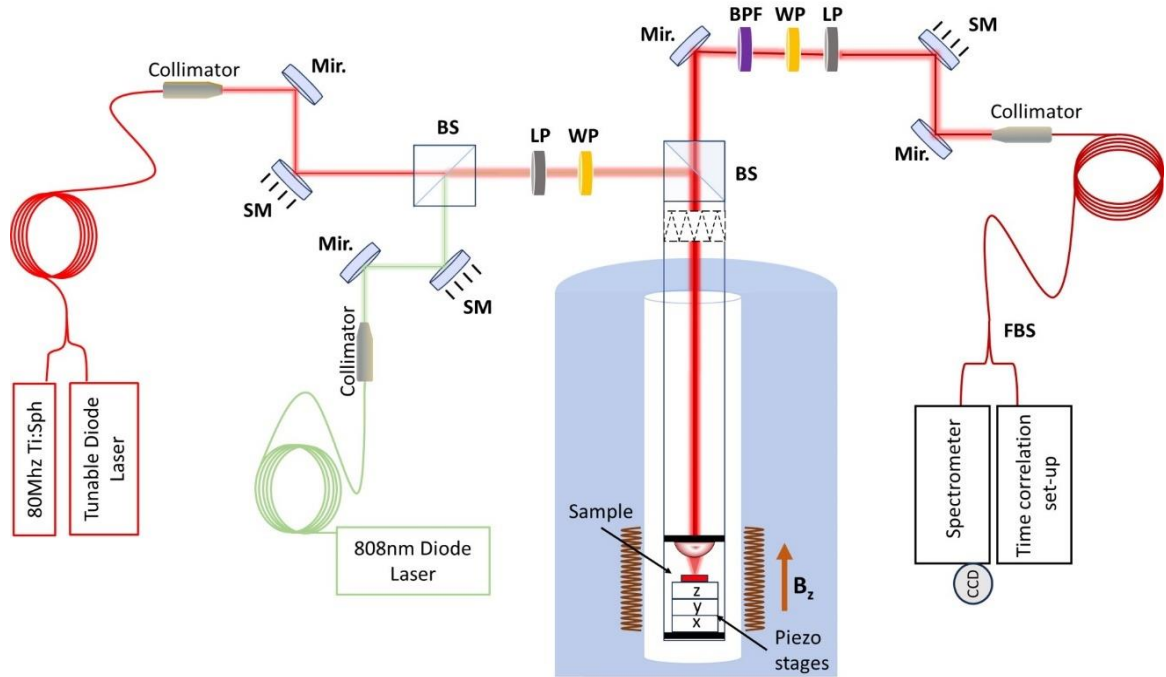

**Figure S1**: Schematic diagram of the experimental setup. SM: scanning mirror; Mir.: dielectric mirror; BS: beam splitter cube; LP: linear polarizer; WP: halfwave plate; BPF: narrow band pass filter (acting as a flip mirror, which is off the path when observing the full spectrum); FBS: fibre beam splitter; CCD: charged coupled device camera. The time correlation setup, used for both HBT and PL decay time measurements, includes two fast superconducting nanowire single-photon detectors (SNSPDs) and a single-photon counting module. The liquid helium cryostat features a sample space surrounded by superconducting coils, enabling magneto-photoluminescence studies with a Faraday-geometry magnetic field applied along the sample axis.

The experimental setup (Figure S1) consists of scanning dielectric mirrors (SM) and dielectric mirrors (Mir.) to direct the excitation and detection paths. On the setup's left-hand side, a linear polarizer (LP) and a half-wave plate (WP), rotate the linear polarisation of excitation laser light to be matched with the angle of linearly polarised outcoupler's mode. On the right-hand side, the combination of LP and WP serves to enhance the collection efficiency of the setup. A narrow band-pass filter (BPF) filters the spectrum by selecting only the bandwidth containing the photoluminescence (PL) emission line under investigation for time-resolved PL and correlation measurements. When in the mode of observing the full spectrum, the BPF is off the path. Two fast superconducting nanowire single-photon detectors (SNSPDs) facilitate time-correlated measurements as parts of a Hanbury and Twist setup. A liquid helium cryostat houses the sample for magneto-photoluminescence studies under a magnetic field perpendicular to the device's plane.

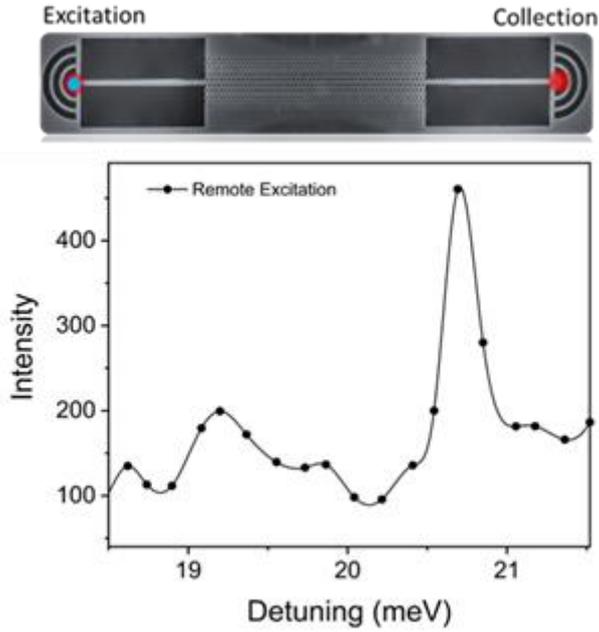

**Figure S2**: PL integrated intensity graph of the exciton line as a function of the CW laser excitation energy detuning from the emission of the QD exciton line in the remote, in-plane excitation experimental scheme.

Figure S2 shows a photoluminescence excitation (PLE) spectrum of QD1 in the vicinity of the p-shell transition. P-shell excitation experiments provided the ability to initialise the circularly polarized optical state efficiently through the relaxation process from p-states to the ground, preserving the spin memory of the initial p-state. Special care was taken during the measurements to maintain the stability of the quasi-resonant excitation energy, as the total diode field (comprising the built-in and applied fields) varied under different excitation conditions (direct and in-plane). This variation was primarily due to the differing excitation power densities of the photogenerated optical field and its contribution to the total diode field. This enabled us to compare the effects of local and remote excitation schemes on directional contrast measurements.

**Section S2**: HBT correlation measurements

Figure S3 shows Hanbury Brown–Twiss (HBT) measurements of the second-order correlation function $g^2(\tau)$ under pulsed excitation on the p-shell absorption line, operating below the saturation regime of the exciton state. The multiphoton emission at zero delay is highly suppressed, dropping to approximately 21% of the adjacent peaks, demonstrating the ability of the QD to act as a single-photon source. The value of $g^2(0)$ is higher than in the CW results in Figure 4c on account of the different pumping conditions. Through autocorrelation measurements, we have shown strong photon antibunching behaviour in both CW and pulsed excitation schemes, validating this QD as a single-photon source suitable for chiral quantum-optical experiments. For pulsed excitation measurements, a weak non-resonant laser was used in addition to the quasi-resonant laser to stabilise charge fluctuations in the quantum dot vicinity, as the emission lines were broad under pulsed excitation. This helped to improve the signal quality, particularly in the lifetime measurements.

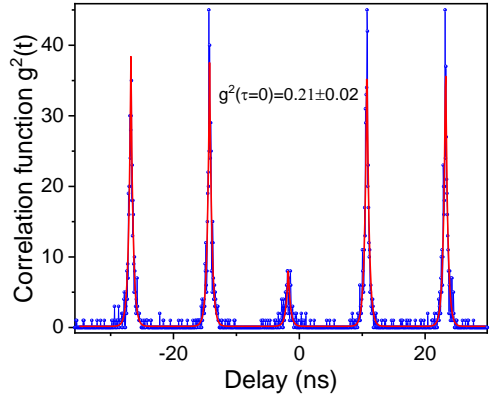

**Figure S3**: Auto-correlation PL histograms without background correction obtained using an HBT setup. The second-order correlation function under pulsed excitation conditions is shown.

**Section S3**: Simulation

The finite-difference time-domain (FDTD) calculations were performed with the *Lumerical* software package and use a simulation region of 80a in length, 16a in width, and 5a in height (where a is the lattice constant of the crystal), with a mesh size of 20 grids per lattice constant. Eight Perfectly Matched Layers (PML) are introduced at the boundary of the simulation region, with an extended thickness of 42 layers applied to the interface perpendicular to the waveguide to eliminate back-reflection and potential divergence issues.

The simulation is conducted on a GaAs/AlGaAs dielectric slab with etched nanoholes patterned as a hexagonal lattice. The following parameters are used in the simulation: refractive index n=3.406 ($\epsilon_r$=11.6), hole radius of the bulk crystal r=0.3a, slab thickness t=0.7a, and lattice constant a=258nm. The radii and positions of the eight innermost nanoholes (perpendicular to the waveguide) are modified for improved slow-light performance based on ref. 21 in the manuscript.

Two linear and orthogonal electric point-dipole sources with a π/2 phase difference are introduced to simulate a circular dipole transition in Faraday geometry, around which a transmission box with a side length of 400 nm is placed. The simulation is set to run over a period of 4000 fs but can be terminated early if the total energy of the electromagnetic field drops below the shut-off threshold (i.e., $10^{-5}$).

The band structure and the mode profile of the glide-plane photonic crystal slab waveguide is calculated using 3D guided-mode expansion method (GME) implemented by *Legume*. There, the eigenmodes of the photonic structures are decomposed and expanded into a basis where the corresponding eigenmode problem is easy to solve (i.e. a homogeneous dielectric slab). Unless otherwise specified, a value of $g_{max}$=4 is used throughout the numerical calculation to improve precision. Here, $g_{max}$ represents the truncation level of reciprocal lattice vectors used in the simulations. Guided modes are expanded on an infinite set of basis vectors in the reciprocal lattice, where truncation is applied to balance simulation time with accuracy. The simulation region is comprised of one single period of the glide-plane waveguide, that is, a rectangle with 16a in height (perpendicular to the waveguide) and a in width (parallel to the waveguide). Periodic boundary conditions are applied to interfaces perpendicular to the GaAs membrane.

**Section S4**: Experimental determination of the slow-light region

The device's slow-light region was found by comparing the PL transmission spectrum with band structure simulations, as detailed in ref. 21. The PL transmission spectrum was recorded by exciting one end of the waveguide through the outcoupler with an 808 nm laser. The laser excites the QD ensemble and wetting layer around the outcoupler, and the PL serves as an internal light source. Subsequently, photons that propagate through the waveguide were collected from the outcoupler at the opposite end. The resulting spectrum (Figure S4a) exhibited a distinct long-wavelength cut-off, dictated by the glide-plane adaptor band gap rather than the glide-plane waveguide, which lacks a cut-off due to the absence of a band gap.

By simulating the band structure of the glide-plane adaptor and aligning it with the observed cut-off, we determined the parameters for the glide-plane waveguide and its slow-light band edge. For example, the slow-light edge for device 1, where QD1 is embedded, was deduced from PL spectrum at approximately 910 nm, consistent with the corresponding transmission spectra (Figure S4). Further details, including specific design parameters, scaling factors, and methodologies, are provided in ref. 21.

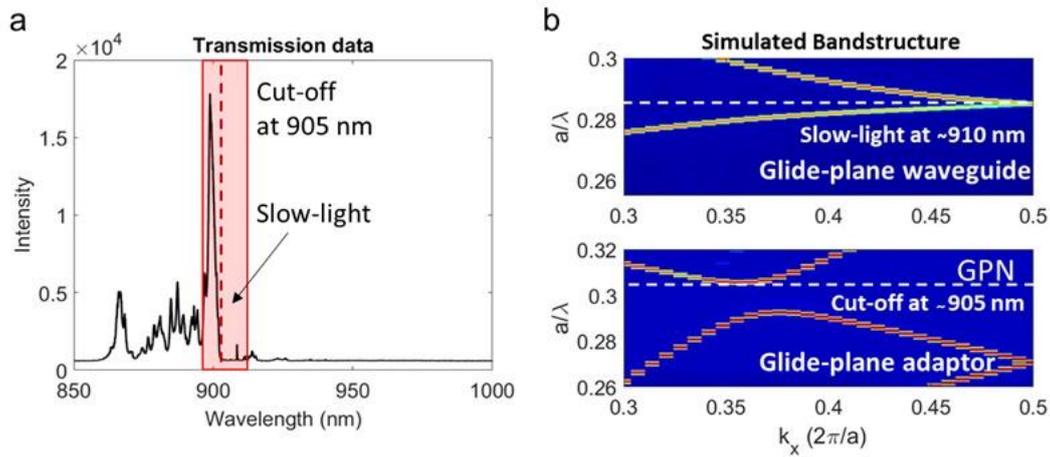

**Figure S4**: (a) High-power PL transmission spectrum recorded from Device 1. (b) Simulated band structure for the glide-plane waveguide (top) and the glide-plane adaptor (bottom). The lattice constant for the glide-plane slow-light section is a=258 nm, with an extended lattice constant of a′=1.07a for the glide-plane adaptor.